\documentclass{emulateapj}
\usepackage{epsfig}
\usepackage{amsmath}
\usepackage{natbib}
\usepackage{color}

\newcommand{\mean}[1]{\langle#1\rangle}

\begin{document}

\title{Was the Sun Born in a Massive Cluster?}
\author{Donald Dukes and Mark R. Krumholz}
\affil{Department of Astronomy \& Astrophysics, University of California, Santa Cruz, CA 95064}

\begin{abstract}
A number of authors have argued that the Sun must have been born in a cluster of no more than several thousand stars, on the basis that, in a larger cluster, close encounters between the Sun and other stars would have truncated the outer Solar System or excited the outer planets into eccentric orbits. However, this dynamical limit is in tension with meteoritic evidence that the Solar System was exposed to a nearby supernova during or shortly after its formation; a  several thousand star cluster is much too small to produce a massive star whose lifetime is short enough to have provided the enrichment. In this paper we revisit the dynamical limit in the light of improved observations of the properties of young clusters. We use a series of scattering simulations to measure the velocity-dependent cross-section for disruption of the outer Solar System by stellar encounters, and use this cross-section to compute the probability of a disruptive encounter as a function of birth cluster properties. We find that, contrary to prior work, the probability of disruption is small regardless of the cluster mass, and that it actually decreases rather than increases with cluster mass. Our results differ from prior work for three main reasons: (1) unlike in most previous work, we compute a velocity-dependent cross section and properly integrate over the cluster mass-dependent velocity distribution of incoming stars; (2) we recognize that $\sim 90\%$ of clusters have lifetimes of a few crossing times, rather than the $10-100$ Myr adopted in many earlier models; and (3) following recent observations, we adopt a mass-independent surface density for embedded clusters, rather than a mass-independent radius as assumed many earlier papers. Our results remove the tension between the dynamical limit and the meteoritic evidence, and suggest that the Sun was born in a massive cluster. A corollary to this result is that close encounters in the Sun's birth cluster are highly unlikely to truncate the Kuiper Belt unless the Sun was born in one of the unusual clusters that survived for tens of Myr. However, we find that encounters could plausibly produce highly eccentric KBOs such as Sedna.
\end{abstract}


\section{Introduction}
It is very likely that our Sun, like most stars, was born as part of a star cluster. The birth cluster environment is a self-gravitating system of gas and dust that collapses at a multitude of points to create its membership of stars. This process is inefficient, converting only 10\% to 30\% of the mass to stars while leaving 70\% to 90\% as gas \citep{LadaLada-2003}. As long as the gas remains, the stars are gravitationally bound and orbit chaotically throughout the cluster. This motion continues until the winds and ionizing radiation of the newborn stars blow away the remaining gas. At that point the cluster is no longer self-gravitating, it disperses, and the stars are free to move about the Galaxy.
\par
By examining the state of our Solar System as it exists today, we can place limits on the birth cluster's properties (i.e. mass, surface density, radius, etc.) and thus have a better understanding of the environment in which our Solar System was born.
\par
Meteorites provide one line of evidence.  Samples taken from meteorites that solidified not long after the Solar System formed show daughter products of multiple radioactive isotopes.  These radioactive isotopes most likely originated from stellar nucleosynthesis, and their presence suggests that there was a nearby supernova that enriched the Solar Nebula \citep{ThraneEtAl-2006, Wadhwa-2007, AllenEtAl-2007, ConnellyEtAl-2008, DupratTatischeff-2008, Adams-2010}.  Such early exposure is likely only if the birth cluster includes a star $\geq 25 M_{\odot}$ in mass. This consideration suggests a birth cluster with at least 1000 members \citep{Adams-2010}. A priori such a large cluster is not unlikely.  The young cluster mass function is $\frac{dN}{dM}\propto M^{-2}$ over the range from $\sim 10^{1}-10^{6}M_{\odot}$ \citep{LadaLada-2003,Chandar-2009,FallEtAl-2010}, which implies that $\sim\frac{1}{2}-\frac{2}{3}$ of stars are born in clusters larger than this.\footnote{We do caution here that the observations do not completely establish the existence of a mass function $dN/dM\propto M^{-2}$ over this full mass range. This powerlaw is seen in both the Galactic sample of \citet{LadaLada-2003} and the extragalactic samples of \citet{Chandar-2009} and \citet{FallEtAl-2010}. However, the \citet{LadaLada-2003} sample only includes clusters within 2 kpc of the Sun, the most massive of which is $\sim 10^3$ $M_\odot$. The \citet{Chandar-2009} and \citet{FallEtAl-2010} samples only go down to a mass of $\sim 10^4$ $M_\odot$. Thus the data do not rule out the possibility that there is a break in the mass function in the range $10^3 - 10^4$ $M_\odot$. However, it would be quite a coincidence for this discontinuity to coincide exactly with the mass range where the observations are incomplete. A simpler hypothesis is that $dN/dM \propto M^{-2}$ over the full mass range.}
\par
On the other hand, several authors have attempted to obtain upper limits on the birth cluster size by considering the effects of close encounters with other stars on the young Solar System. Close encounters would perturb the Solar Nebula during its formation. Any such perturbation must still permit the formation of a Solar System like the one we see today, with eight planets that all lie close to the same plane (inclination angles $\leq$ 3.5$^{\circ}$) and have small eccentricities (less than 0.2, or less than 0.09 is we exclude Mercury, whose eccentricity is pumped by perturbations from other planets). 
\par
\citet{AdamsLaughlin-2001} determined an upper bound on the birth cluster size by performing Monte Carlo scattering simulations of close encounters between the Solar System and passing stars. They found that the maximum number of stars the birth cluster could have had without causing a Jovian planet to be excited to an eccentricity greater than 0.1 is $2200\pm1100$.  However, observations of the properties of young, embedded star clusters were at the time quite limited, and they were forced to make a variety of assumptions about cluster properties.  First, the relative velocities of the stars in their simulations were chosen randomly from a Maxwellian distribution with a standard deviation of 1 km/s; thus they had very few simulations done with higher velocities, and the velocity distribution was implicitly assumed to be independent of the cluster mass.  Second, they assumed that the cluster's lifetime, and thus the time the Sun was exposed to close encounters, was on the order of 100 cluster crossing times.  Lastly, they assumed that the cluster's radius was fixed and independent of its mass.  This last assumption would lead a high mass cluster to have very high surface densities.
\par
More recent observations have shown that most of these assumptions are not satisfied in typical embedded clusters. Compiling observations of embedded clusters from \citet{ShirleyEtAl-2003}, \citet{FaundezEtAl-2003}, and \cite{FontaniEtAl-2005}, \citet{FallEtAl-2010} find that embedded clusters form a sequence of roughly constant surface density, not constant radius. Similarly, current observations favor cluster lifetimes  closer to one to four crossing times \citep{Elmegreen-2000,HartmannEtAl-2001,LadaLada-2003,TanEtAl-2006, JeffriesEtAl-2011,ReggianiEtAl-2011}, not 100. After $\sim 10$ Myr, 90\% of the stars born in clusters have dispersed into the field, and by $\sim 100$ Myr roughly 99\% have done so \citep{AllenEtAl-2007,FallEtAl-2009, ChandarEtAl-2010}. We do not know the lifetime of the Sun's birth cluster, but in obtaining limits on its size it is important to consider the possibility both that the lifetime was very short, and that it was tens of Myr, rather than assuming the latter as was done in most previous work.
\par
In this paper we show that retaining the same theoretical framework as \citet{AdamsLaughlin-2001}, but using these more recent observations to set the properties of the embedded clusters leads to a dramatic relaxation of the \citet{AdamsLaughlin-2001} limit on the size of the Sun's birth cluster.  In turn, this greatly eases the tension between the meteoritic and dynamical constraints.  In particular,  the lifetime of a $25 \ M_{\odot}$ star before it goes supernova is $\sim7.54$ Myr, longer than most clusters survive. A 75 $M_{\odot}$ star, however, has a lifetime of $\sim3.64$ Myr. \citet{Adams-2010} finds that to have a reasonable probability of producing such a star, a birth cluster of membership size $>10^{4}$ is needed. Again, such a large cluster is not a priori unlikely, given the observed young cluster mass function.
\par
Close encounters within the Sun's birth cluster have also been theorized to be responsible for other aspects of the Solar System's present-day architecture, particularly the drastic drop in object density in the Kuiper Belt at $\sim$45AU, and the Kuiper Belt object Sedna.  Both of these phenomena are difficult to understand in our current picture of Solar System formation.  The existence of Neptune, with its size and orbit, suggests that the surface density of the Solar Nebula should have been high enough to produce Kuiper Belt Objects (KBOs) out to distances greater than 50AU with a gradually declining object density.  An unperturbed Solar Nebula, which tends to produce objects with low eccentricities, is hard to reconcile with Sedna's eccentricity of 0.84 and semi-major axis of 542 AU.  A close encounter with another star could produce a disk truncation \citep{IdaEtAl-2000,KobayashiIda-2001} and scatter objects into high eccentricity orbits to produce Sedna \citep{BrasserEtAl-2006}. 
\par
Close encounters could also affect other putative objects orbiting the Sun: the proposed companion brown dwarf Nemesis \citep{Raup-1984, Whitmire-1984} or the more recently proposed Tyche \citep{Matese-1999,MateseEtAl-2005,MateseWhitmire-2011}. Tyche has been proposed to have an orbit just inside the Inner Oort Clound (a semi-major axis of $\sim$5,000 AU) with a mass of 1-4 Juipter Masses.  Given the low probability of capturing such a companion \citep{Tohline-2002}, if either Nemesis or Tyche exists, they must have formed with the Sun.  To date, there has been no research on whether Nemesis nor Tyche could survive the dynamics of the birth cluster.

\par
In this paper we address all of these issues. We have constructed simulations similar those of \citet{AdamsLaughlin-2001}. We use these together with updated observational data on the properties of embedded clusters to determine an upper limit of our birth cluster's membership size and the effects of a close encounter on both the truncation on the Kuiper Belt and production of a Sedna-like object.  In addition, we will use this data to determine the likelihood of Tyche staying bound to the Sun during its time within the birth cluster.  \citet{AdamsEtAl-2006} have undertaken similar calculations, but survey a much more limited part of parameter space.  In particular, they do not consider clusters of more than 10$^{3}$ stars. In Section 2 we will look at the simulation and method used in our experiment. In Section 3 we will then analyze the results of the simulation and finally in Section 4 we will make our conclusions.

\section{Methods}
To determine an upper limit on the birth cluster membership size, we want to determine the probability of a disruptive encounter as a function of cluster surface density and mass.  By disruptive event, we mean a close encounter in which any of the Jovian planets are excited to an eccentricity greater than 0.1.  In the process, we will also determine whether encounters with passing stars could be responsible for truncating the Kuiper Belt at $\sim$50AU, for producing a Sedna-like object, or for stripping any distant brown dwarf companion.
\subsection{Determining Probability}
To obtain the probability of an event occurring, we first perform numerical scattering simulations to determine the velocity-dependent cross-section for that event, $\langle\sigma\rangle$. We then calculate the average cross section times velocity, $\langle\sigma v\rangle$. We defer a description of our simulations to section 2.2. Simulations of star clusters suggest that the distribution of encounter velocities within a embedded stellar cluster is Maxwellian \citep{ProszkowAdams-2009}, so
\begin{equation}
	\label{eq:sigma}
	\langle\sigma v\rangle=\frac{1}{\sigma^{3}_{v}}\sqrt{\frac{2}{\pi}}\int_{-\infty}^{\infty}\langle\sigma\rangle v^{3}{\it e}^\frac{-v^{2}}{2\sigma^{2}_{v}}dv
\end{equation}
where $\sigma_{v}$ is the standard deviation of the velocity distribution.  If we then multiply this by the stellar number density of the cluster, $n$, we obtain the encounter rate,
\begin{equation}
	\label{eq:Gamma}
	\Gamma=n\langle\sigma v\rangle.
\end{equation}
If we multiply $\Gamma$ by the total time the Sun was exposed in the cluster, we will have the expected number of events, $\Lambda$.  This time will be roughly the time for which the cluster survives before mass evaporation leads it to disperse, $t_{life}$.  Thus, we have
\begin{equation}
	\label{eq:Lambda}
	\Lambda=\Gamma t_{life}.
\end{equation}
To compute the overall probability that an event will occur, we will assume that events follow a Poisson distribution, in which case the probability that an event will occur at least once is
\begin{equation}
	\label{eq:PoissonOne}
	P_{event}=1-{\it e}^{-\Lambda}.
\end{equation}

The calculation is slightly more complicated for the Kuiper Belt, where there are many KBOs, and varying numbers can be affected by a single encounter. In our simulations we represent particular radial bins of the Kuiper Belt with a number $N_t$ of test particles. We use our simulation to determine $\langle\sigma\rangle_{k}$, the velocity-dependent cross section for an event (e.g.\ excitation or ejection from the Solar System) to affect $k$ of these $N_t$ test KBOs. Using equation \ref{eq:sigma}, we can then calculate $\langle\sigma v\rangle_{k}$ for each value of $k$. The rate at which the KBO population undergoes events is then given by 
\begin{equation}
	\label{eq:kGamma}
	\Gamma=n\sum_{k=1}^{N_t}\frac{k}{N_t}\langle\sigma v\rangle_{k},
\end{equation}
where $n$ is the number density determined in equation \ref{eq:NumberDensity}. Once $\Gamma$ has been determined, we  use equation (\ref{eq:Lambda}) to compute $\Lambda$. The expected fraction of KBOs that undergo a certain even is then
\begin{equation}
	\label{eq:expectedfraction}
	\langle f\rangle=1-e^{-\Lambda}.
\end{equation}

The initial conditions of the birth cluster are unknown to us, so we cannot directly compute all of these quantities.  However, we can estimate them in terms of the cluster surface density, $\Sigma_{c}$, and cluster mass, $M_{c}$. For a spherical cluster, the number density, velocity dispersion, and crossing time are
\begin{equation}
	\label{eq:NumberDensity}
	n=\frac{3\pi^{1/2}\Sigma^{3/2}_{c}}{4\overline{m}M_{c}^{1/2}} 
	 = 700 M_4^{-1/2}\Sigma_{-1}^{3/2}\mbox{ pc}^{-3}
\end{equation}
\begin{equation}
	\begin{split}
	\label{eq:VelocityDispersion}
	\sigma_{v}=\left(\frac{3}{5}\alpha_{vir}\right)^{1/2}G^{1/2}(\pi				\Sigma_{c}M_{c})^{1/4} \\
	 = 3.2 \alpha_{\rm vir}^{1/2} M_4^{1/4} \Sigma_{-1}^{1/4}\mbox{ km 		s}^{-1}
	 \end{split}
\end{equation}
\begin{equation}
	\label{eq:CrossingTime}
	t_{cross} =  \frac{M_{c}^{1/4}}{G^{1/2}(\pi\Sigma_{c})^{3/4}}
	= 0.62 M_{4}^{1/4} \Sigma_{-1}^{-3/4}\mbox{ Myr},
\end{equation}

\begin{figure*}
    \plotone{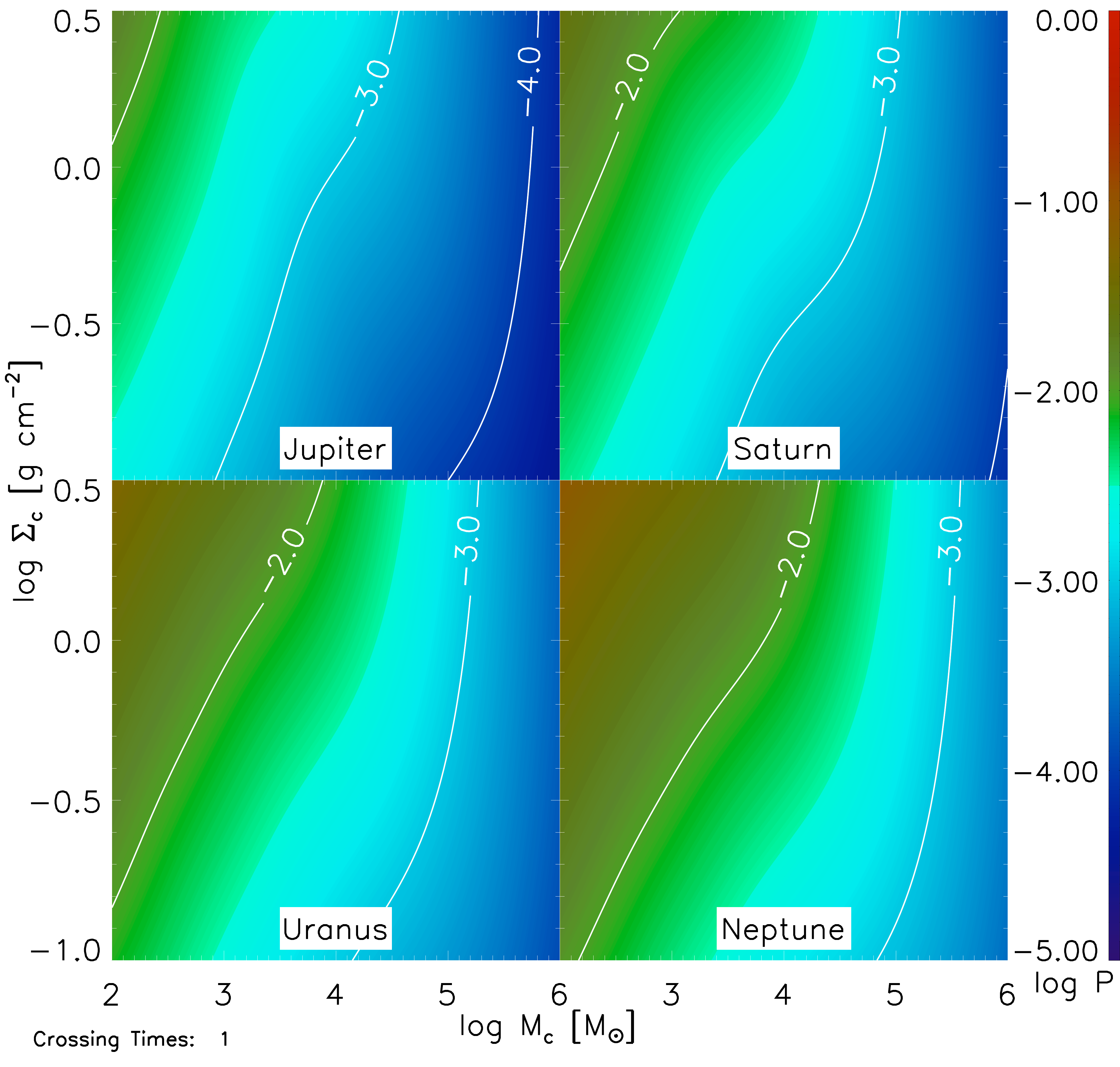}
    \caption{The log of the probability of a close encounter exciting each Jovian planet's eccentricity to greater than 0.1, in one crossing time, as a function of cluster mass M$_{c}$ and surface density $\Sigma_{c}$.
    \label{fig:Jovian}}
\end{figure*}

\begin{figure*}
    \plotone{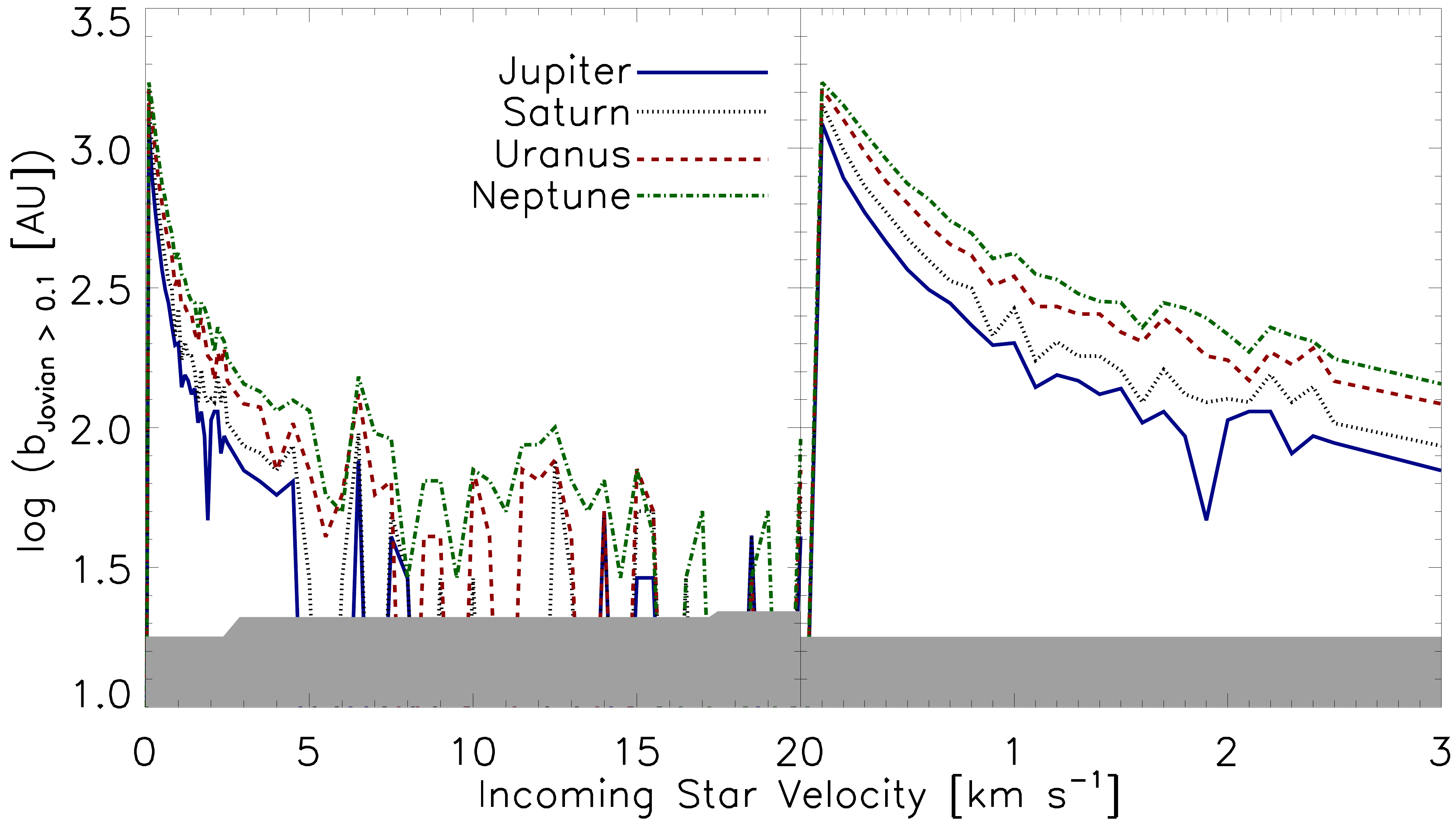}
    \caption{The effective impact parameter, $b=\sqrt{\langle\sigma \rangle_{v} / \pi}$ for exciting a Jovian to eccentricity greater than 0.1 vs.\ velocity of incoming star. They gray band indicates, given then number of trials at a given velocity, the smallest value of $b$ we can measure with 99\% confidence. The increase in the minimum $b$ with velocity is a result of ours having somewhat more simulations at low than at high velocity. The right panel shows a zoomed in portion of the full graph shown on the left.
    \label{fig:JovImpact}}
\end{figure*}

\begin{figure*}
    \plotone{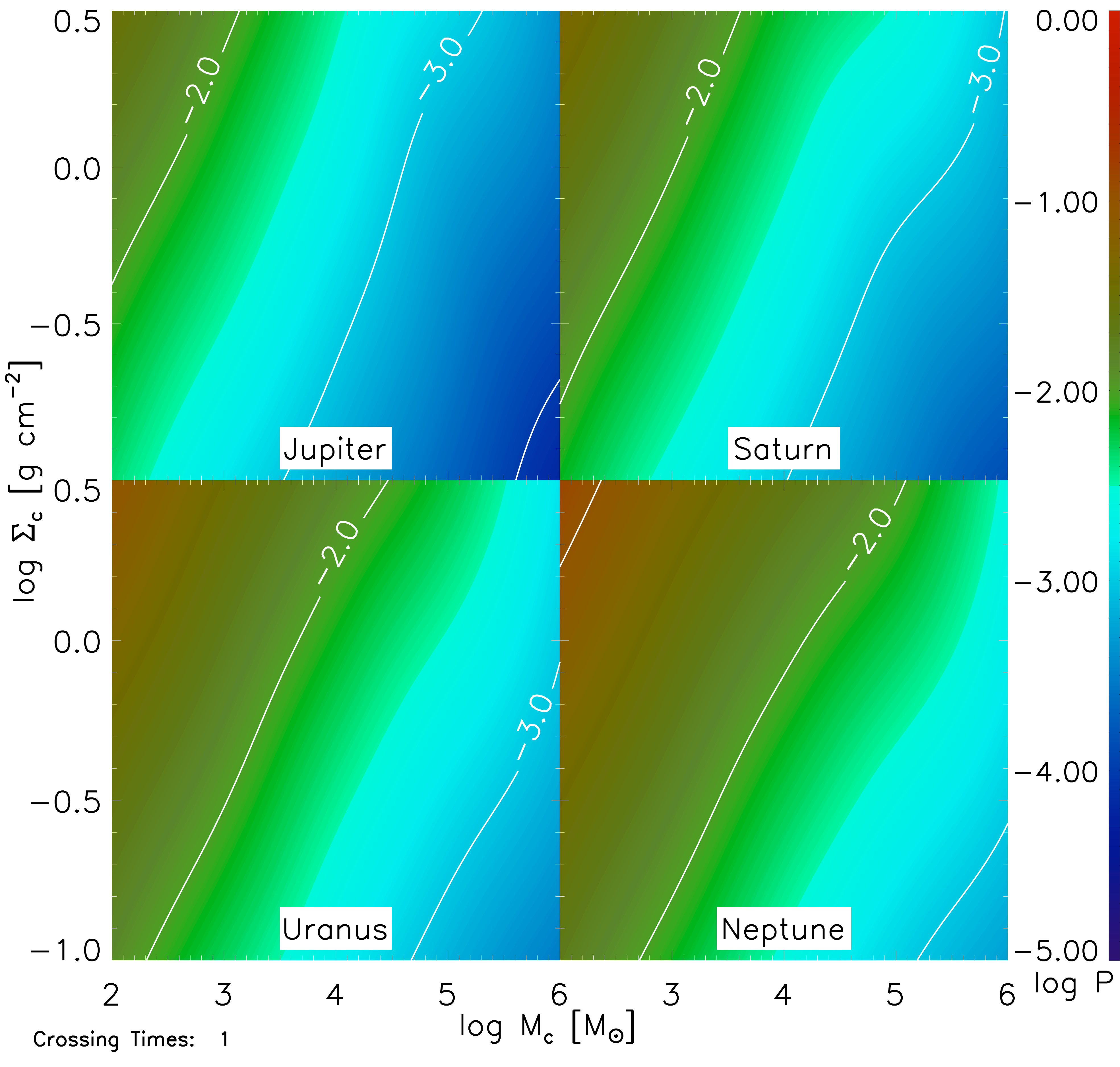}
    \caption{The log of the probability of a close encounter exciting each Jovian planet's eccentricity to greater than 0.1, in one crossing time, as a function of cluster mass M$_{c}$ and surface density $\Sigma_{c}$ in a sub-viral cluster ($\alpha_{vir}=$1/3).
    \label{fig:JovianVir}}
\end{figure*}

where $\alpha_{vir}$ is the virial parameter, $\overline{m}$ is the mean stellar mass, which we take to be 0.2$M_{\odot}$ \citep{Chabrier-2005} for our calculations, $M_{4}=M_{c}/10^{4}$ $M_{\odot}$, and $\Sigma_{-1}=\Sigma_{c}/0.1$ g cm$^{-2}$. Equations (\ref{eq:NumberDensity}) and (\ref{eq:VelocityDispersion}) provide estimates of the number density and velocity dispersion that we can use in our estimates. The relationship between the cluster lifetime $t_{\rm life}$ and the crossing time, however, is somewhat uncertain. As discussed in the Introduction, most clusters disperse rapidly, although exactly how rapidly is debated. For example, the best studied cluster containing O stars is the Orion Nebula Cluster ($M_4 = 0.46$, $\Sigma_{-1} = 1.2$ -- \citealt{Hillenbrand98a}), for which $t_{\rm cross} = 0.44$ Myr. Estimates of its age spread range from $\sim 1-3$ Myr \citep[e.g.][]{ReggianiEtAl-2011, JeffriesEtAl-2011}, corresponding to $\sim 1 - 6 t_{\rm cross}$, depending on the technique used to estimate protostellar ages and similar systematic uncertainties. On the other hand, $\sim 10\%$ of clusters survive for more than 10 Myr. We therefore consider two possible scenarios for the lifetime. The more likely one is that the Sun was born in one of the rapidly-dispersing clusters, and in this case we take $t_{\rm life} = t_{\rm cross}$; the true lifetime may be a factor of a few larger, but, as we will see below, this factor of a few is relatively insignificant. The other possibility is that the Sun was born in the minority of clusters with a longer lifetime. To represent this case, we take $t_{\rm life} = 30$ Myr, typical of this class.

Finally, note that, in the rapidly-dispersing cluster scenario, for massive and high surface density clusters the cluster lifetime may be significantly less that the main sequence lifetime even of a very massive star. This is certainly the case in the ONC, for example, where star formation is complete and the cluster is likely in the process of dispersing, but $\Theta^1$ Ori C is still alive. This is not necessarily a problem for enrichment, however. Even for a dispersing cluster, the relatively low velocity dispersion of the stars implies that stars will not spread that far before the most massive stars go supernova. 

\subsection{Simulations}
The simulations we use to measure $\langle\sigma\rangle$ consist of the Solar System and the scattering star system.  The Solar System includes the Sun, the four Jovian planets, Tyche and 320 KBOs all within the same plane.  We only place Tyche in the simulation because the semi-major axis of Nemesis is so large that we cannot reasonably evaluate it behavior with scattering simulations. Instead, a full (and much more expensive) N-body simulation of the entire birth cluster would be required.  We treat the Jovian planets and KBOs as massless to conserve computing power.  This means that we are unable to model secular processes produced by planet-planet interactions. However, since such processes generally take much longer than that $\sim 0.01$ Myr covered by our simulations, we could not model these in any event. We consider Tyche masses of zero, 0.02 $M_{\odot}$ and 0.08 $M_{\odot}$.  We set the orbital properties of the Jovian planets to their present-day values and we randomize the initial orbital phases. We give Tyche a semi-major axis of 1,000 AU\footnote{While this is closer than the 5000 AU proposed by \citet{MateseEtAl-2005} and \citet{MateseWhitmire-2011}, the use of a similar but smaller distance greatly reduces the computational cost by allowing us to consider a smaller range of incoming star impact parameters.}, an eccentricity randomly between 0.0 and 0.4, and a random argument of periastron. We set up the Kuiper Belt up such that all objects are placed in 32 concentric circles ranging from 35 AU to 500 AU, in 15 AU intervals.  Each circle has ten objects equally spaced in angle, and all objects have a zero eccentricity.  For our statistical analysis, we then divide the KBOs into eight distance bins (35-80AU, 95-140AU, 155-200AU, 215-260AU, 275-320AU, 335-380AU, 395-440AU and 455-500AU), each bin containing four concentric circles. 
\par

The incoming star systems consist of either a solo star or a binary star system.  We first select the mass of the primary incoming star by drawing from a \citet{Chabrier-2005} IMF.  We then determine if the primary has a companion.  We set the probability of there being a companion to
\begin{equation}
	\label{eq:binary}
	f_{binary} = \left\{
	\begin{array}{lr}
     		0.2 & : m\;<\;0.5\;M_{\odot} \\
       		0.2+0.8\left( \frac{m-0.5}{1.5} \right) & : 0.5\;M_{\odot}\;\leq\;m\;\leq 2.0\;M_{\odot} \\
       		1.0 & : m\;>\;2.0\;M_{\odot}
     	\end{array}
   	\right.	
\end{equation}
based on a rough fit to the mass-dependent companion fraction reported by \citet{Lada-2006}. If a companion exists, we also select its mass from the IMF, rejecting and redrawing if the mass exceeds the primary's mass.  Finally, we select the orbital period of the binary by drawing from the distribution observed for field G star binaries \citep{DuquennoyMayor-1991}
\begin{equation}
	\label{eq:tau}
	p(\log\tau )=\frac{1}{\sqrt{2\pi \sigma_{\log\tau}^{2}}}e^{\frac{-(\log\tau -\overline{\log\tau})^{2}}{2\sigma_{\log\tau}^{2}}}
\end{equation}
where $\tau$ is the period in days, $\sigma_{\log\tau}$=2.3, and $\overline{\log\tau}=4.8$.
\par
We determine the impact parameter of the encounter, $b$, randomly with probability proportional to $b$, with a maximum, $b_{max}$, of 2000 AU.  We randomize the relative orientations of the incoming star, the Solar ecliptic, and the binary orbital plane.  Finally, for each incoming system, we run the simulation for a series of relative velocities between the system and the Sun.  We use velocities of $v$=0.1-2.5 km/s at intervals of 0.1 km/s and $v$=3.0-20.0 km/s at intervals of 0.5km/s.  Over 2.1 million runs were preformed in total.
\\ \\ 
\section{Results}

\begin{figure}
    \plotone{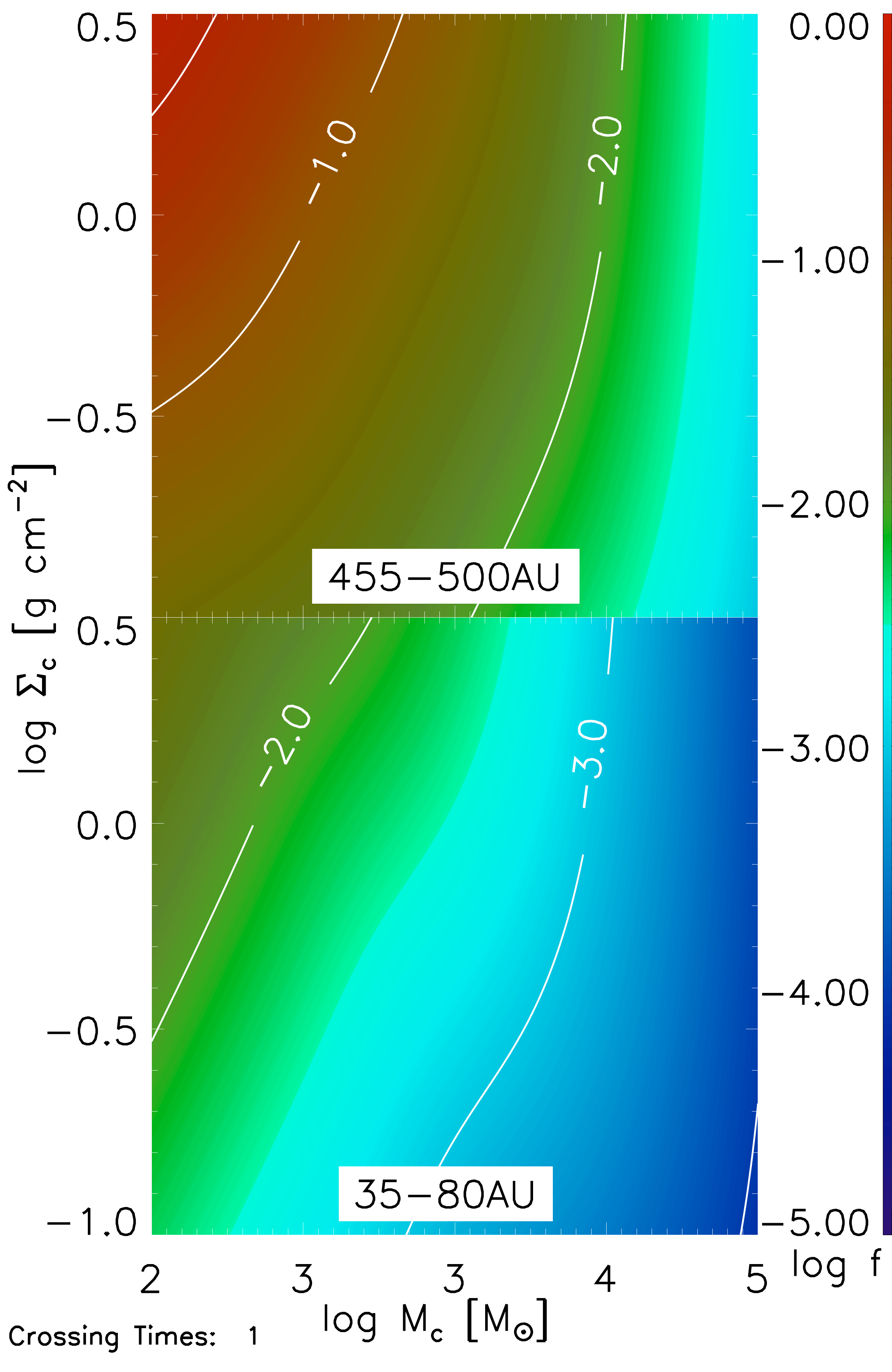}
    \caption{The log of the expected fraction of KBOs that will become unbound in one cluster crossing time in a given distance bin.  In all cases, the inner bin and the outer bin represent the extreme values.}
    \label{fig:KBOLost01CT}
\end{figure}

\begin{figure}
    \plotone{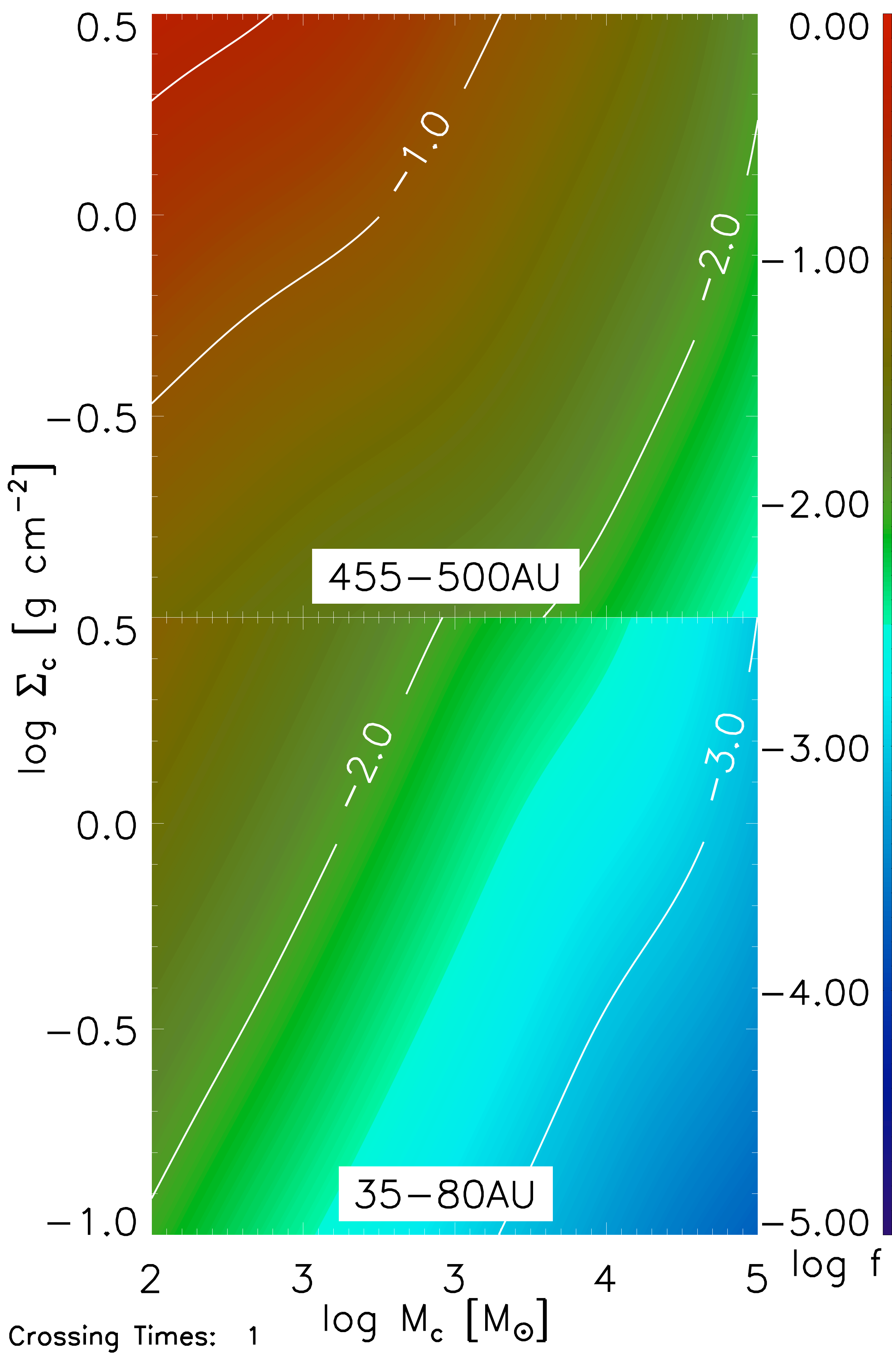}
    \caption{The log of the expected fraction of KBOs that will become unbound in a sub-viral cluster ($\alpha_{vir}=$1/3) in one cluster crossing time in a given distance bin.  In all cases, the inner bin and the outer bin where the extreme values.}
    \label{fig:KBOLostVir01CT}
\end{figure}

\begin{figure}
    \plotone{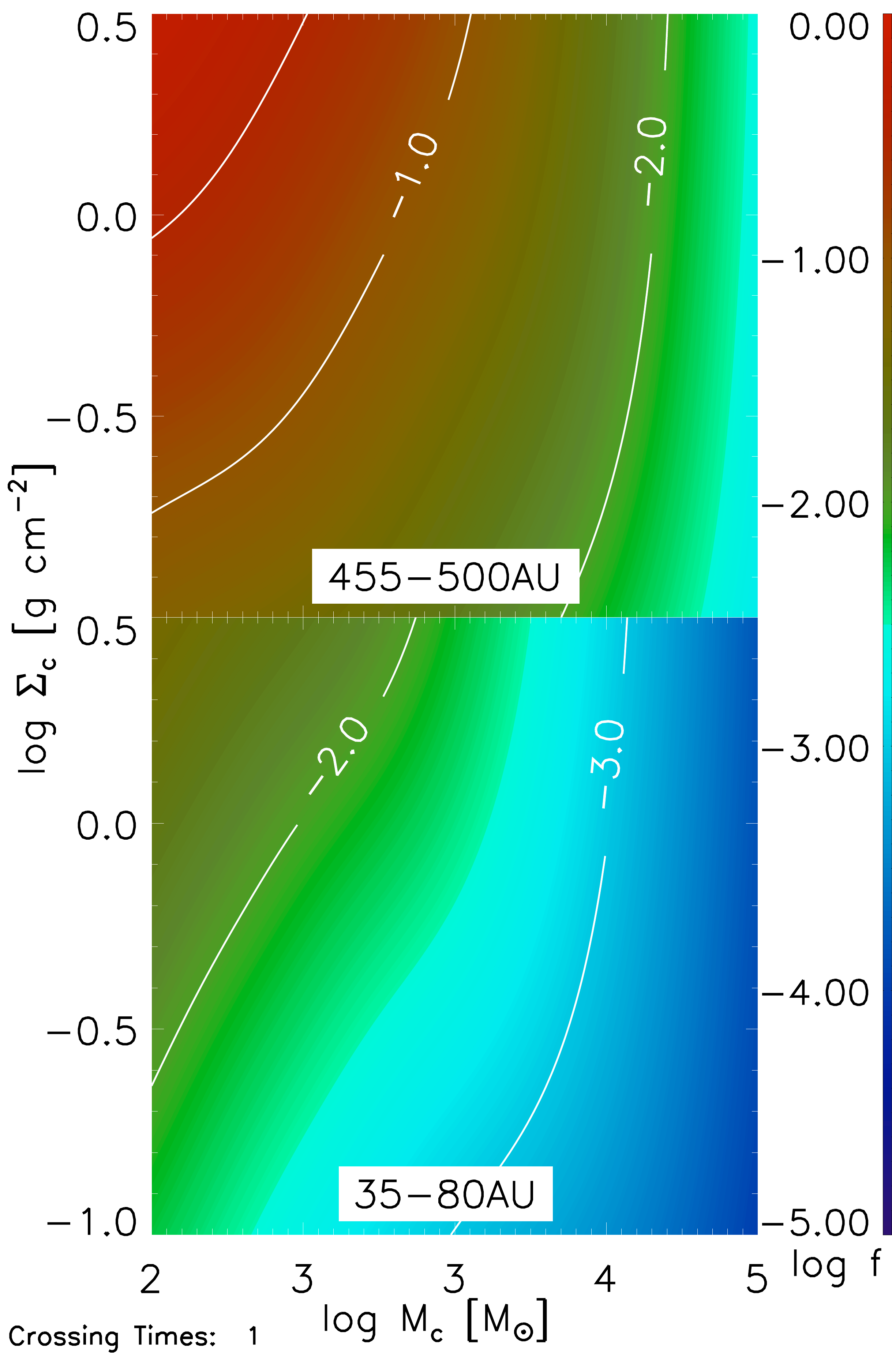}
    \caption{The log of the expected fraction of KBOs that will be excited to an eccentricity greater than 0.5 but less that 1.0 in one cluster crossing time in a given distance bin.  In all cases, the inner bin and the outer bin represent the extreme values.}
    \label{fig:KBOExcited01CT}
\end{figure}

\begin{figure}
    \plotone{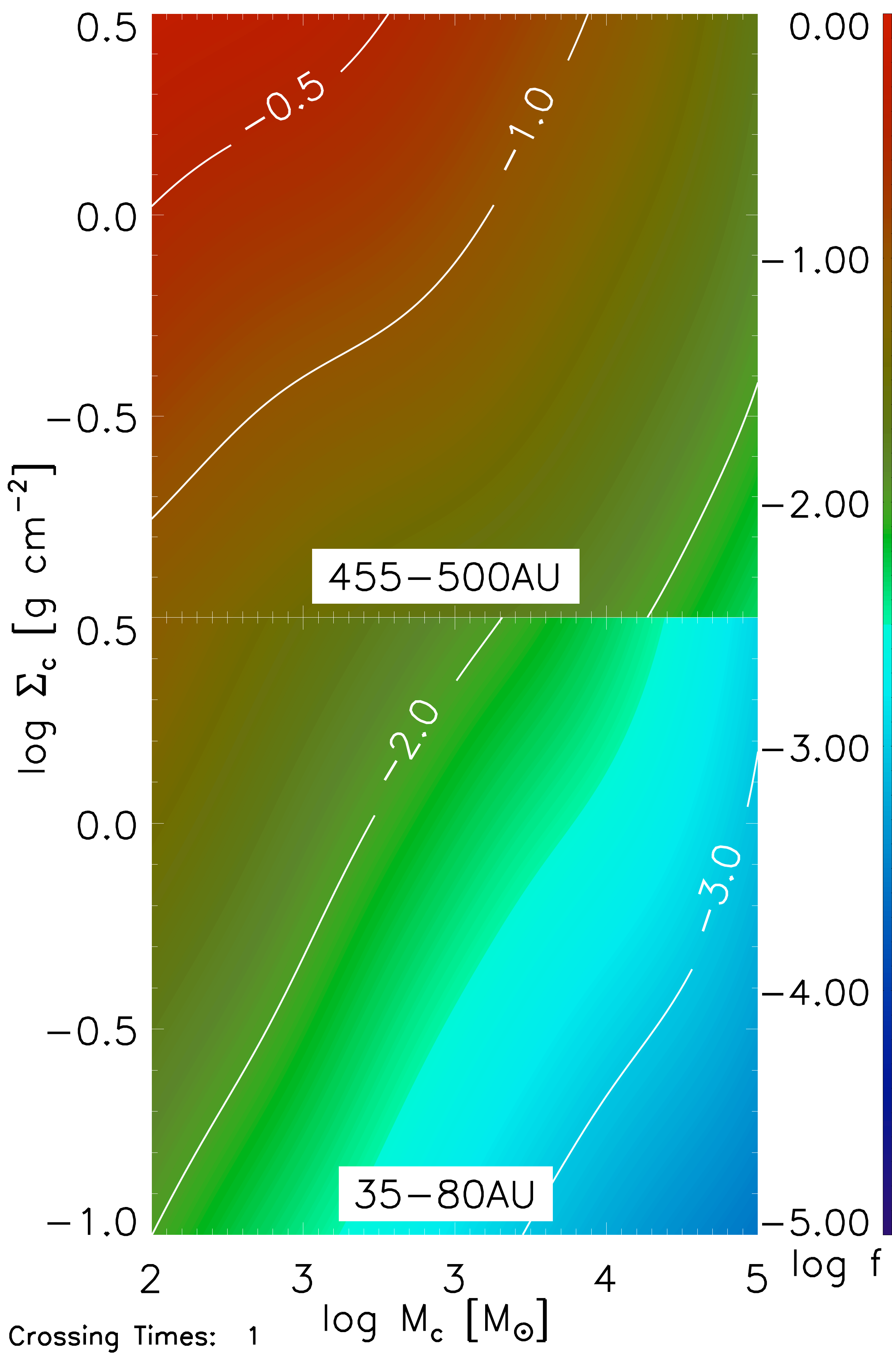}
    \caption{The log of the expected fraction of KBOs that will be excited to an eccentricity greater than 0.5 but less that 1.0 in a sub-viral cluster ($\alpha_{vir}=$1/3) in one cluster crossing time in a given distance bin.  In all cases, the inner bin and the outer bin where the extreme values.}
       \label{fig:KBOExcitedVir01CT}
\end{figure}

\begin{figure}
\epsscale{1.0}
    \plotone{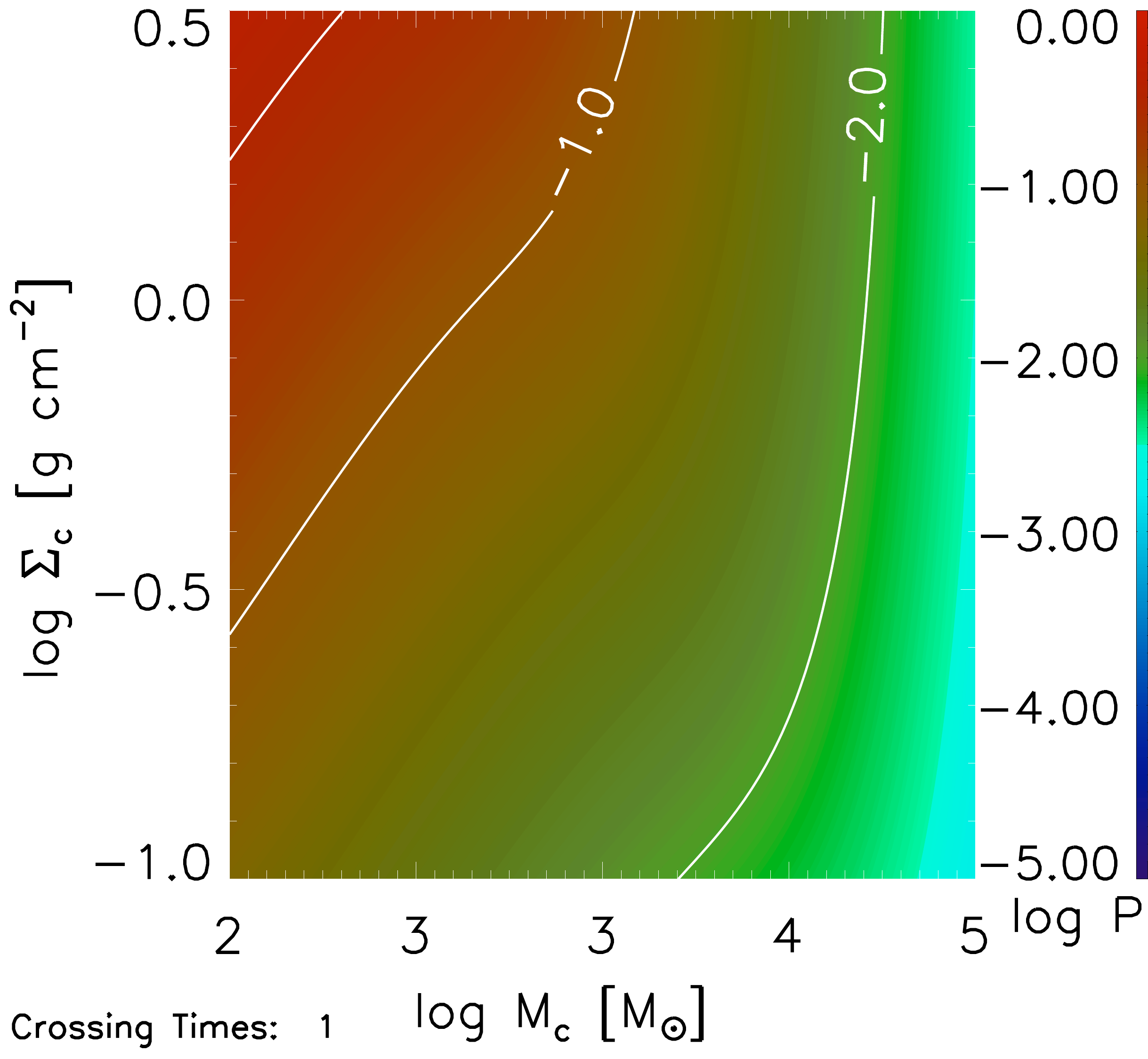}
    \caption{The log of the probability that, in one crossing time, Tyche becomes unbound.}
    \label{fig:Tyche}
\end{figure}

\begin{figure}
    \plotone{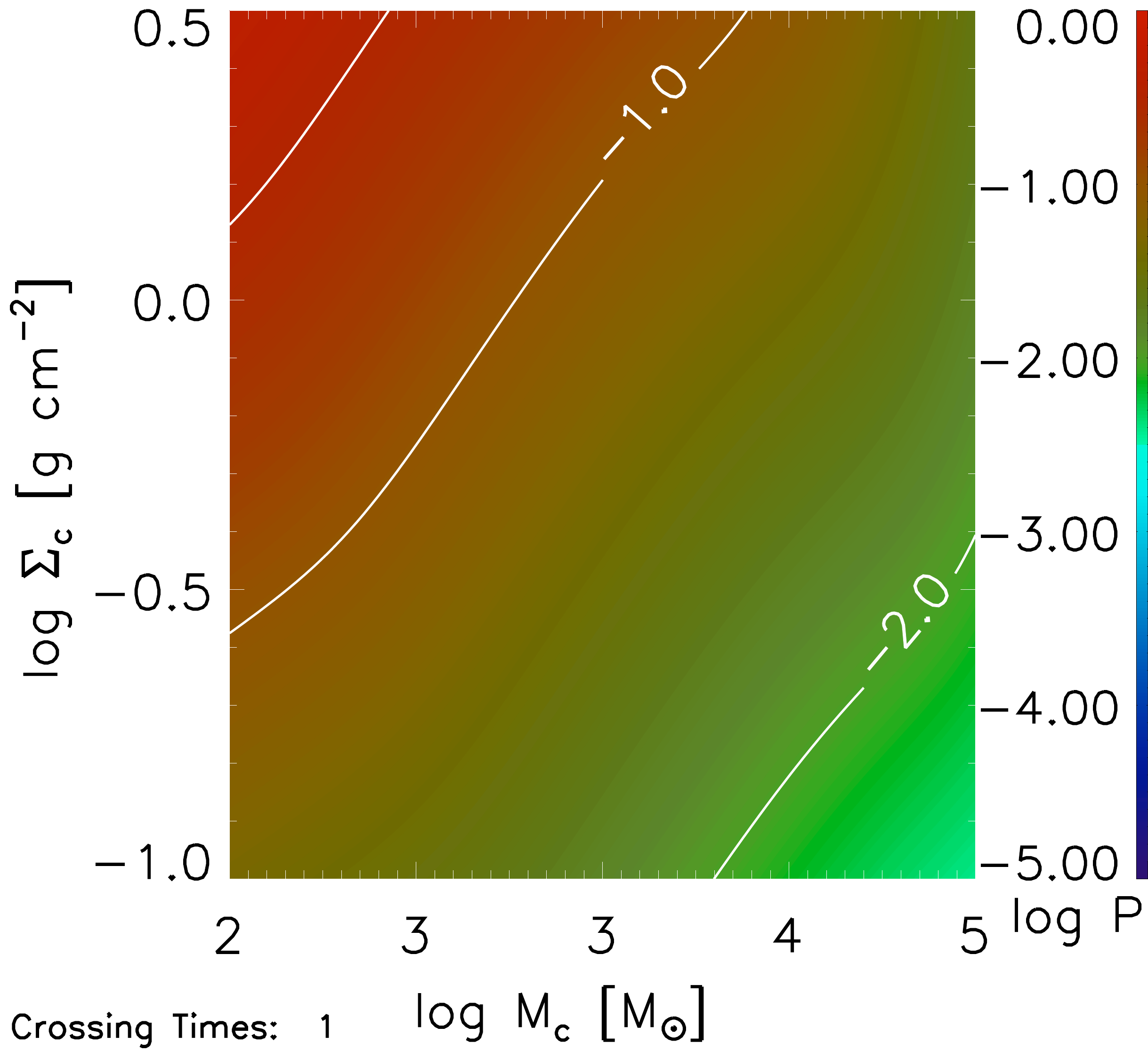}
    \caption{The log of the probability that, in one crossing time, Tyche becomes unbound in a sub-viral cluster ($\alpha_{vir}=$1/3).
    \label{fig:TycheVir}
    }
\end{figure}

In Sections \ref{sec:jovian} -- \ref{sec:tyche}, we first examine the effects of encounters on the Jovian planets, on the Kuiper Belt, and on Tyche in the more likely scenario where the Sun was born in a cluster with a lifetime $t_{\rm life} = t_{\rm cross}$. In Section \ref{sec:longlife} we examine how these results change in the scenario where the Sun's parent cluster was one of the $\sim 10\%$ that reach ages of $t_{\rm life} = 30$ Myr.

\subsection{Effects of Close Encounters on the Jovian Planets}
\label{sec:jovian}

\begin{figure*}
    \epsscale{0.8}
    \plotone{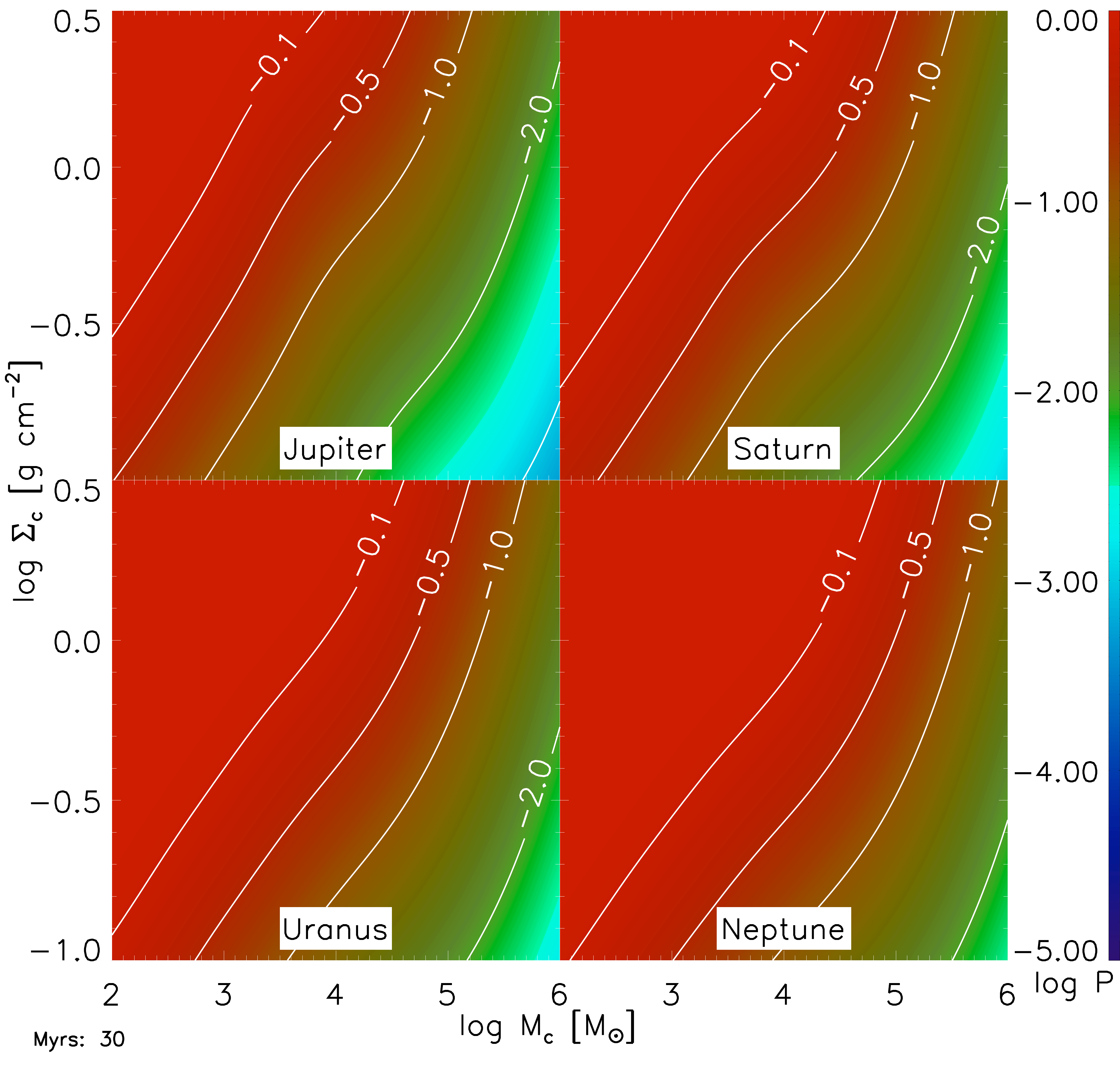}
    \caption{The log of the probability of a close encounter exciting each Jovian planet's eccentricity to greater than 0.1 as a function of cluster mass M$_{c}$ and surface density $\Sigma_{c}$ with cluster lifetime of 30 Myr.}
    \label{fig:JovianMyr30}
\end{figure*}

\begin{figure}
    \plotone{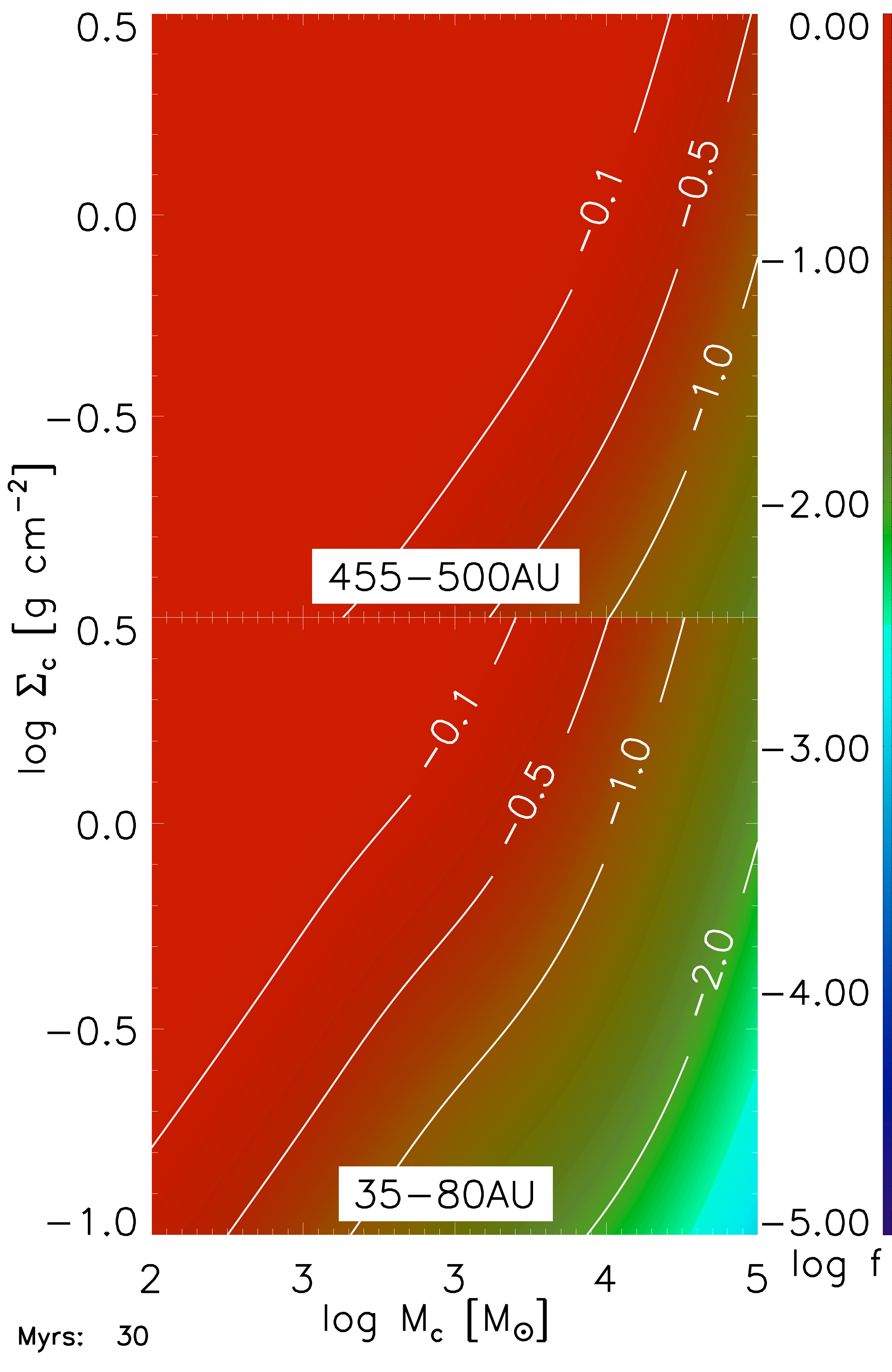}
    \caption{The log of the expected fraction of KBOs that will become unbound with cluster lifetime of 30 Myr in a given distance bin.  In all cases, the inner bin and the outer bin represent the extreme values.}
    \label{fig:KBOLostMyr30}
\end{figure}

\begin{figure}
    \plotone{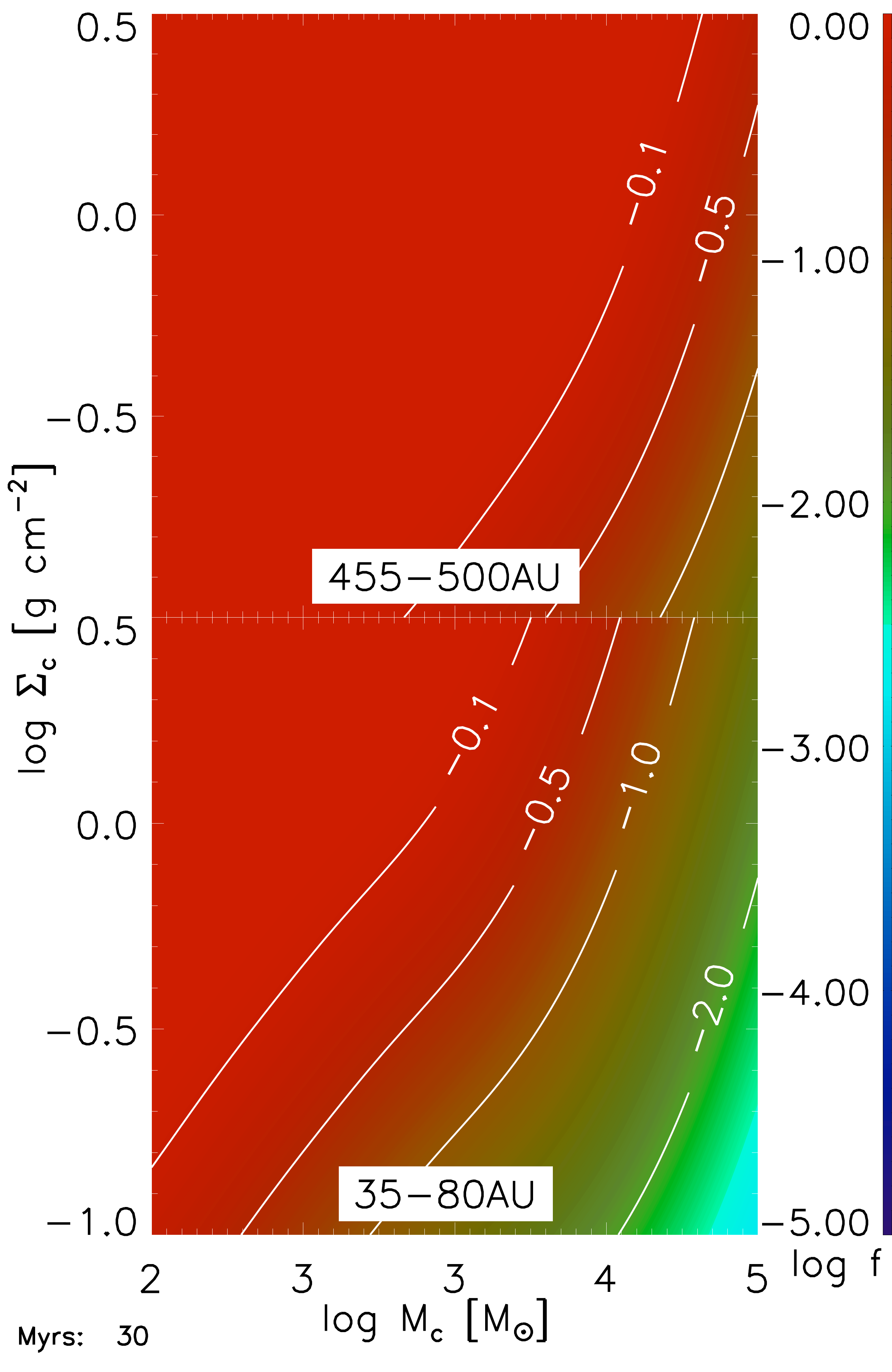}
    \caption{The log of the expected fraction of KBOs that will be excited to an eccentricity greater than 0.5 but less that 1.0 with cluster lifetime of 30 Myr in a given distance bin.  In all cases, the inner bin and the outer bin represent the extreme values.}
    \label{fig:KBOExcitedMyr30}
\end{figure}

Figure \ref{fig:Jovian} show the probabilities of each Jovian planet being excited to an eccentricity greater than 0.1 as a function of $M_{c}$ and $\Sigma_{c}$ in a fully relaxed embedded cluster (i.e. $\alpha_{vir}=$5/3 corresponding to $\sigma_{v}=\sqrt{(GM_{c})/R_{c}}$).  We use a range of 10$^{-1}$-10$^{0.5}$ g cm$^{-2}$ for $\Sigma_{c}$ because this covers the range of embedded cluster properties complied by \citet{FallEtAl-2010}.  Similarly, we use masses from 10$^{2}$-10$^{6}$M$_{\odot}$ because this covers the range of cluster masses seen in nearby galaxies and in the Milky Way \citep{LadaLada-2003,FallEtAl-2009,ChandarEtAl-2010}. Looking at the four graphs, we see that the probability is very low for all four planets in almost all combinations of $\Sigma_{c}$ and $M_{c}$ within the plausible range of mass and surface density.  The probability is less than 1\% for all four Jovian planets when the mass of the cluster is greater than 10$^{4}M_{\odot}$.  Neptune, the furthest planet from the Sun and thus the most likely to be excited, has excitation probabilities $<$1\% over most of parameter space, and reaches a high of 8.7\% at $M_{c}=10^2M_{\odot}$, $\Sigma_{c}=10^{0.5}$ g cm$^{-2}$. 
\par    
A curious feature of Fig.\ \ref{fig:Jovian} is that as the mass of the cluster increases, the probability of an event decreases. To understand this, we first must realize that the number of encounters is independent of $M_{c}$ at fixed $\Sigma_{c}$. This is because $\Lambda \propto \frac{n_{c}t_{cross}} {\sigma_{v}^{3}}M_{c} \propto M^{0}_{c}$.  Next, looking at equation (\ref{eq:VelocityDispersion}) we see that as $M_{c}$ increases, so does the velocity dispersion.  Figure \ref{fig:JovImpact} shows the velocity dependence of the cross section obtained though our simulation. We see that, at low velocities around 1 km/s, the cross section is high, comparable to the values obtained by \citet{AdamsLaughlin-2001}. However, at slightly higher velocities the cross section dramatically decreases, dropping under (200 AU)$^{2}$ at around 3 km/s.  Thus, at higher $M_{c}$ the velocity dispersion increases yet the number of encounters does not. This explains the result in Fig.\ \ref{fig:Jovian}: higher mass clusters are less likely to increase a Jovian planet's eccentricity because they produce no more encounters (at fixed $\Sigma_{c}$), but reduce the effective cross section per encounter. 
\par
Recent observational and theoretical evidence indicates that young embedded clusters can be sub-virial \citep{FureszEtAl2008,TobinEtAl-2009,OffnerEtAl-2009} and thus we make the same calculations for a cluster with a virial parameter that is one fifth the value of a fully relaxed cluster, $\alpha_{vir}=$1/3.  Figure \ref{fig:JovianVir} shows probability of exciting a Jovian planet to an eccentricity greater than 0.1 in a sub-virial cluster.  We see that the probability is higher than that of a relaxed cluster, which makes sense due to the lower velocity dispersions,  but the overall probability is still small.  Again, Neptune is the likeliest to be excited with $\sim$15\% probability at the high $\Sigma_{c}$, low $M_{c}$ extreme, but as in a relaxed cluster, a majority of parameter space produces probability below 1\% for all of the Jovian planets.  This agrees with \citet{ProszkowAdams-2009} who show that the interaction rate in a sub-virial cluster is greater than that in a virialized one, but not by much.

\subsection{Effects of Close Encounters on the Kuiper Belt}
We next check if close encounters could truncate the Kuiper Belt or excite an object to a Sedna-like orbit.  To determine if the a close encounter could truncate the Kuiper Belt, we compute the expected fraction of the KBOs in each of our eight initial distance bins (see Section 2.1 for derivation).  The resulting expected fractions in the 35-80AU and 455-500AU bins, which represent the extremes of our initial Kuiper Belt, are show in Fig.\ \ref{fig:KBOLost01CT} for relaxed clusters and Fig.\ \ref{fig:KBOLostVir01CT} for sub-virial clusters.  We find that the expected fraction of KBOs stripped in the 35-80AU bin is less than 1\% for most of parameter space, only going above 1\% in the high $\Sigma_{c}$, low $M_{c}$ extreme, where the maximum is $\sim$4\%.  The expected fraction of KBOs stripped in the 455-500 AU bin reaches 88\% in the high $\Sigma_{c}$, low $M_{c}$ extreme. As we move away from this extreme, the probability decreases drastically. This appears to suggest that a birth cluster with $\Sigma_{c}$ and low $M_{c}$ could explain the truncation of the Kuiper Belt. However, comparing Fig.\ \ref{fig:KBOLost01CT} with Fig.\ \ref{fig:Jovian} we note that any cluster capable of destroying the outer Kuiper Belt while leaving the inner Kuiper Belt undisturbed would also disrupt Neptune's orbit. There is no part of parameter space where the Kuiper Belt is truncated but Neptune is not also perturbed. This is most likely because no single encounter can truncate the Kuiper Belt. Instead, multiple close stellar passes would be required, and the number required is sufficiently large as to make it extremely likely that one of those encounters would be close enough to perturb Neptune.
\par
To see if a close encounter could produce a Sedna-like object, we compute the expected fraction of the KBOs excited to an eccentricity between 0.5 and 1.0 in a given distance bin. We show the result in Fig. \ref{fig:KBOExcited01CT} for relaxed clusters and Fig.\ \ref{fig:KBOExcitedVir01CT} for sub-virial clusters.  Again, we see that the greatest probability is found in the high $\Sigma_{c}$, low $M_{c}$ extreme, with a significant drop-off as we move away from this part of the parameter space.  Since Sedna has a semi-major axis of 542 AU and there is only one such object known, we focus on the 455-500 AU distance bin, and we only need a relatively small probability. We see reasonable probabilities exist over much of our parameter space.  We conclude that it is plausible that a close encounter could produce a Sedna-like object.

\subsection{Effects of Close Encounters on Tyche}
\label{sec:tyche}
To determine if Tyche could survive the birth cluster, we compute the cross section for Tyche becoming unbound.  We first check to see if there was any difference in the cross-section depending upon which Tyche mass was used (see section 2.2).  Using a chi-square test, we find that our measurements of the number of Tyche losses as a function of incoming star velocity at our three different Tyche masses are consistent with each other to $\sim$98\% confidence, and thus there is no reason to reject the hypothesis that the probability of Tyche becoming unbound is independent of it mass. Thus, we use a combination of data from all three masses.  Looking at Fig. \ref{fig:Tyche}, as before, the highest probability, $\sim$44\%, is in the high $\Sigma_{c}$, low $M_{c}$ extreme, with the probability decreasing as we move away from this extreme. The probability is $\sim$10\% for moderate $\Sigma_{c}$ and M$_{c}$. Given this, there is a small likelihood that Tyche would be stripped away. However, recall that we place Tyche at 1,000AU, and while it is theorized to have a semi-major axis of $\sim$5,000 AU. Thus the probability of stripping likely exceeds our estimate by a considerable factor. Even with this increased, though, stripping seems unlikely in a cluster with mass $>10^{4}M_{\odot}$.

\subsection{Encounter Effects in Long-Lived Clusters}
\label{sec:longlife}

Finally, we consider the possibility that the natal cluster did not disperse at the point where its gas was expelled, and instead lived on as an open cluster. As discussed above, about 10\% of embedded clusters experience this fate. In this scenario, the Sun would be exposed to many more close encounters. To study the effects of this scenario, we use $t_{life}=30$ Myr in equation \ref{eq:Lambda} instead of using equation \ref{eq:CrossingTime} and repeat our analysis from the previous sections.
\par
Figure \ref{fig:JovianMyr30} shows the probability of the Jovian planets being excited to an eccentricity greater than 0.1 in such a cluster. As we found for the more typical short-lived cluster population, the probability of disruption is smallest for clusters of high mass and low surface density. Not surprisingly, excitation of the Jovians is much more likely in a long-lived cluster, and the only part of parameter space that has low probabilities is the high mass, low surface density extreme. Outside this regime all the Jovian planets are likely excited to an eccentricity greater than 0.1. Thus we conclude that the low eccentricities of the planets are consistent with the long-lived cluster scenario, but only if the cluster in question was both quite massive and of low surface density.
\par
Figures \ref{fig:KBOLostMyr30} and \ref{fig:KBOExcitedMyr30} show the expected fraction of KBOs lost and excited, respectively, in a long-lived cluster. We see that the expected fraction lost is very high for KBOs in all distance bins except in clusters of very high mass and low surface density. There is a fairly narrow strip of parameter space that allows the outer Kuiper Belt to be destroyed but leaves the inner Kuiper Belt relatively untouched. However, comparing to Figure \ref{fig:JovianMyr30}, we see that in this parameter regime it is also very likely that Neptune would be driven to high eccentricity. The underlying reason is that same as for short-lived clusters: it takes many encounters to truncate the outer Kuiper Belt, and any cluster capable of producing that many encounters is also likely to supply one close enough to disturb Neptune. The behavior of Sedna is also similar to that seen in the case of short-lived clusters: there is a reasonably large part of parameter space where the Jovian planets would not be disturbed, but there is a reasonable change of exciting a distant KBO to Sedna-like eccentricities.

\section{Discussion}
Our conclusion that encounters in the Sun's natal cluster are likely to have very little effect on the architecture of the Solar System, even for very massive clusters, is at odds with much previous work, which \citet{Adams-2010} summarizes to conclude that dynamical arguments imply that the Sun's birth cluster could not have contained more than several thousand stars. This conclusion, however, is based on work that assumed the cluster lifetime was much greater than current observational evidence shows, that the surface density of a cluster varies linearly with the cluster's mass, and that only simulated encounters at fairly low velocities. Each of these assumptions increases the probability of disruption encounters.  For example  \citet{LaughlinAdams-1998} estimated $\mean{\sigma}$=(230 AU)$^2$ for disruption of Jovian planets due to interactions with binary stars in the birth cluster through scattering simulations similar to ours.  To obtain this value, they randomly chose the velocity of the incoming star from a normal distribution with $\sigma=1$ km s$^{-1}$.  From this, they determined the rate of disruptive encounters is $\sim$0.13 per 100 Myr. \citet{AdamsLaughlin-2001} used the same velocity distribution and obtained similar results. We can immediately identify the effects of changing observational data which lead us to different conclusions.  First, consulting equation (\ref{eq:VelocityDispersion}), we note that a cluster of mass $10^{4}M_{\odot}$ and surface density 1 g cm$^{-2}$ will have a velocity dispersion of 7 km s$^{-1}$ (if it is virialized) or 3 km s$^{-1}$ (for $\alpha_{vir}=1/5$).  As a result most encounters will be at significantly higher velocities than Laughlin \& Adams assumed; consulting Fig.\ \ref{fig:JovImpact}, we see that this will reduce the cross section by roughly an order of magnitude.  The earlier work was also based on a lifetime of 100 Myr, while modern observations tells us that only $\sim$1\% of stars remain in a cluster for that long, and only $\sim 10\%$ survive to even 10 Myr \citep{LadaLada-2003,FallEtAl-2009,ChandarEtAl-2010}. Again, switching to modern values leads to a 1-2 order of magnitude drop in the disruption probability.  Together the decrease in cross section and in exposure time are sufficient to explain why we find no dynamical limit on the size of the Sun's parent cluster.
\par
Changing the assumed cluster velocity dispersion and lifetime also explains why our conclusions on the Kuiper Belt differs from that of earlier research. \citet{LestradeEtAl-2011} concluded that it was possible to truncate a debris disk around a main sequence star in an open cluster by a close encounter. To reach this conclusion, however, they assumed parabolic, coplanar encounters over a 100 Myr cluster lifetime, compared to our observationally-favored scenarios of mostly hyperbolic, non-coplanar encounters over a $\sim$1 Myr cluster dissolution timescale. \citet{Jimenez-Torres-2011} created a simulation that was able to produce Sedna-like objects but assumed incoming star velocities of 1, 2 and 3 km s$^{-1}$, closest approach distances of 200 AU and cluster lifetimes of 100 Myr. We concur with these authors that it is possible to truncate the Kuiper Belt in a cluster that lives 30 Myr, and that it may be possible to produce Sedna even in a shorter lived cluster. That the Sun might have been born in a long-lived cluster is by no means impossible, since $\sim 10\%$ of stars are. We simply point out that such a scenario is not typical. Moreover, we find that, regardless of the cluster lifetime, it is very difficult to truncate the outer Kuiper Belt without simultaneously disrupting the orbit of Neptune, a problem not considered in earlier work.

Some other authors who have used the current observational data about cluster lifetime and velocities have partially anticipated the results presented here.  \citet{BonnellEtAl-2001} argue that planet formation is unaffected in open clusters, and that it was possible for a the Sun to be born within a cluster whose density is greater than 10$^3$ stars pc$^{-3}$ if the cluster dissolves within $\sim$10 Myr. Using a cluster membership size of 1000 and a cluster lifetime of 10 Myr, \citet{AdamsEtAl-2006} showed that typical stellar passes are not close enough to appreciably enhance the eccentricity of Neptune. Their main conclusion was that planet-forming disks and newly formed solar systems generally survive their aggregates with little disruption.  \citet{SpurzemEtAl-2009} constructed a N-body simulation of the Orion Nebula Cluster, an environment containing massive stars similar to the hypothetical progenitors of the supernova(e) that produce the isotopes we see in meteorites.  Their conclusion was that the critical threshold for the survival of wider orbit  planets (those similar to our Jovian planets) is a cluster lieftime less than 10$^8$ yr. 
\par
We note that \citet{BrasserEtAl-2006} have shown through N-body simulations that both close encounters and cluster tidal effects have a reasonable probability of creating a Sedna-like object if the object begins at a semi-major axis $\ge$300AU, consistent with our conclusions (see Fig.\ \ref{fig:KBOExcited01CT} and Fig.\ \ref{fig:KBOExcitedVir01CT}). They hypothesize a two-step formation process for Sedna in which the outward migration of Neptune scatters a KBO into an orbit with large semi-major axis, and the object is scattered again by passing stars.  Our work is not inconsistent with this scenario.

\section{Conclusion}
In this paper we revisit the tension between meteoritic and dynamical constraints on the size of the Solar System's parent star cluster.  The meteorite evidence suggests that a supernova deposited short-lived radioactive isotopes near the forming Solar System.  However, based on dynamical arguments about the likelihood of a close encounter disrupting the outer Solar System, \citet{AdamsLaughlin-2001} and \citet{Adams-2010} argue that the birth cluster could not have had no more than several thousand members. The probability of having a supernova within the range needed in a cluster of that size is only about 2\%. Given that we have only a single example (the Sun) of a star system that simultaneously has supernova enrichment and outer planets in circular orbits, this isn't necessarily impossible. It could simply be that our Solar System is quite unusual. However, we find that no such conclusion is warranted. We show that repeating the calculations of \citet{AdamsLaughlin-2001} but with embedded cluster properties drawn from recent observations, we can update the upper limit on birth cluster size to values where supernova pollution is much more probable.
\par
Indeed, we show that as the mass of the birth cluster increases, the relative velocities of the constituents increases and the effective impact parameters for Solar System-disrupting events decreases, yet the number of expected encounters is constant. Our result differs from previous work because we are using current observational evidence that suggests that the cluster's surface density is independent of its mass, and its lifetime is much shorter than previously believed.  The removal of the upper bound from dynamics allows the cluster size to be large enough to be able to produce a massive star which can go supernova to seed the Solar System with the observed short lived radioactive isotopes.  Indeed, at cluster sizes of 10$^4$-10$^5$ stars, multiple supernovae are likely, allowing each supernova to be further away and exposing the Solar System to less mass loss.
\par
A close encounter within the birth cluster has also been proposed to explain the existence of Sedna. We find that this is within reason, showing that it is possible to produce a small fraction of KBOs at $\sim$500 AU with eccentricity between 0.5 and 1. However, we find that it is highly unlikely that encounters in the birth cluster could be responsible for truncation of the Kuiper Belt.  There is only a small region of parameter space that allows for the destruction of the outer Kuiper Belt while leaving the inner Kuiper Belt intact, and any cluster capable of disrupting the outer Kuiper Belt would also be very likely to excite Neptune to a higher eccentricity than we observe.

Finally, we caution that disruption of planetary orbits by the gravity of passing stars may not be the only limiting factor when it comes to cluster size.  \citet{Adams-2010} argues that in a cluster of more than $\sim 10^4$ stars, the ultraviolet radiation produced by massive stars would photoevaporate the Sun's protoplanetary disk, preventing formation of the outer planets. However, this conclusion is highly sensitive to the rate of mass loss due to photoevaporation, which is highly uncertain. The $10^4$ star limit is based on a loss rate taken from \citet{AdamsEtAl-2004}, but \citet{ErcolanoEtAl-2009} show that this rate is uncertain at the order of magnitude level. Moreover, it is not entirely clear that photoevaporation inhibits planet formation; instead, by raising the dust-to-gas ratio, it may trigger gravitational instability \citep{Throop-2005}. Due to these uncertainties, the question of whether photoevaporation might provide a limit on the cluster size remains an open one for future research.

\acknowledgements
We thank G.~Laughlin for extensive discussions and comments on the manuscript. MRK is supported by the Alfred P.~Sloan Foundation, the National Science Foundation through grants AST-0807739 and CAREER-0955300, and NASA through Astrophysics Theory and Fundamental Physics Grant NNX09AK31G and a Chandra Space Telescope Grant.

\bibliography{ClusterDynamics}

\end{document}